\documentstyle[psfig,epsf,conf-X]{article}
\newcommand{\mincir}{\raise -2.truept\hbox{\rlap{\hbox{$\sim$}}\raise5.truept
\hbox{$<$}\ }}
\newcommand{\magcir}{\raise -2.truept\hbox{\rlap{\hbox{$\sim$}}\raise5.truept
\hbox{$>$}\ }}
\newcommand{\siml}{\raise -2.truept\hbox{\rlap{\hbox{$\sim$}}\raise5.truept
\hbox{$<$}\ }}
\newcommand{\simg}{\raise -2.truept\hbox{\rlap{\hbox{$\sim$}}\raise5.truept
\hbox{$>$}\ }}

\newcommand{\be}{\begin{equation}}
\newcommand{\ee}{\end{equation}}
\newcommand{\ba}{\begin{eqnarray}}
\newcommand{\ea}{\end{eqnarray}}
\newcommand{\brr}{\begin{array}}
\newcommand{\err}{\end{array}}
\newcommand{\bc}{\begin{center}}
\newcommand{\ec}{\end{center}}

\newcommand{\hm}{\,h^{-1}{\rm Mpc}}
\newcommand{\vel}{\,{\rm km\,s^{-1}}}
\newcommand{\fl}{\,{\rm erg\,s^{-1}cm^{-2}}}

\begin{document} 
\small
\heading{%
%
The ROSAT Deep Cluster Survey: Constraints on Cosmology
}
\par\medskip\noindent
\author{%
S. Borgani$^{1,2}$, P. Rosati$^{3}$, R. Della Ceca$^4$,
P. Tozzi$^{5,6}$, C. Norman$^6$ 
}
\address{%
INFN - Sezione di Trieste, c/o Dipartimento di Astronomia, via Tiepolo
11, I-34131 Trieste, Italy
}
\address{%
INFN - Sezione di Perugia, c/o Dipartimento di Fisica, via A. Pascoli,
I-06121 Perugia, Italy
}
\address{%
European Southern Observatory, D-85748 Garching bei M\"unchen, Germany
}
\address{%
Osservatorio Astronomico di Brera, via Brera 28, I-20121 Milano, Italy
}
\address{%
Osservatorio Astronomico di Trieste, via Tiepolo 11, I-34131 Trieste, Italy
}
\address{%
Dept. of Physics and Astronomy, The Johns Hopkins University,
Baltimore MD 21218, USA
}

\begin{abstract}
  We use the ROSAT Deep Cluster Survey (RDCS) with the purpose of
  tracing the evolution of the cluster abundance out to $z\simeq 0.8$
  and constrain cosmological models. We resort to a phenomenological
  prescription to convert masses into $X$--ray fluxes and
  apply a maximum--likelihood approach to the RDCS redshift-- and
  luminosity--distribution. As a main result we find that, even
  changing the shape and the evolution on the $L_{bol}$--$T_X$
  relation within the observational uncertainties, a critical density
  Universe is always excluded at more than $3\sigma$ level. By
  assuming a non--evolving $X$--ray luminosity--temperature relation
  with shape $L_{bol}\propto T_X^3$, it is
  $\Omega_m=0.35^{+0.35}_{-0.25}$ and $\sigma_8=0.76^{+0.38}_{-0.14}$
  ($\Omega_m=0.42^{+0.35}_{-0.27}$ and
  $\sigma_8=0.68^{+0.21}_{-0.12}$) for flat (open) models, while no
  significant constraints are found for the power--spectrum shape
  parameter $\Gamma$. Uncertainties are $3\sigma$ confidence
  levels for three significant fitting parameters.
\end{abstract}
\section{Introduction}
The mass function of local ($z\mincir 0.1$) galaxy clusters has been
used as a stringent constraint for cosmological models. Independent
analyses have shown that $\sigma_8\Omega_m^{\gamma(\Omega_m)}\simeq
0.5$--0.6, where $\Omega_m$ is the density parameter, $\sigma_8$ the
r.m.s. fluctuation amplitude within a sphere of $8\hm$ ($h=H_0/100
\vel$ Mpc$^{-1}$) radius and $\gamma(\Omega_m)\simeq 0.4-0.6$
\cite{ECFH,Geal98}. The increasing availability of $X$--ray
temperatures for distant ($z\magcir 0.3$) clusters is providing a
handle to estimate the density parameter which best reproduces the
evolution of the cluster abundance \cite{ECFH,Don99,Bah98} (see also Henry,
this volume, for a review), A limitation of this approach comes from
the small size of the current samples \cite{VL99}.

An alternative way to trace the evolution of the cluster abundance is
to rely on the luminosity and redshift distribution of $X$-ray
flux--limited cluster samples \cite{SBO98,BRTN99,Real99}.  The
advantage of this approach lies in the availability of large samples,
with well understood selection functions.  As a limitation, however, one
has to face with the uncertain relation between cluster masses and
$X$--ray luminosities.  The ROSAT Deep Cluster Survey (RDCS)
\cite{RDCS} provides a flux--limited complete sample of clusters
identified in the {\sl ROSAT} PSPC archive and including $\magcir 100$
spectroscopically confirmed systems. In the following we will outline
the main results of a comparison between the RDCS sample and the
predictions of cosmological models. The analysis of RDCS for
constraining the evolution of the $X$--ray luminosity function is
contained in a separate paper (Rosati et al., this volume).

\begin{figure}
\epsfxsize 11.5truecm
\hspace{1.5truecm}\epsffile{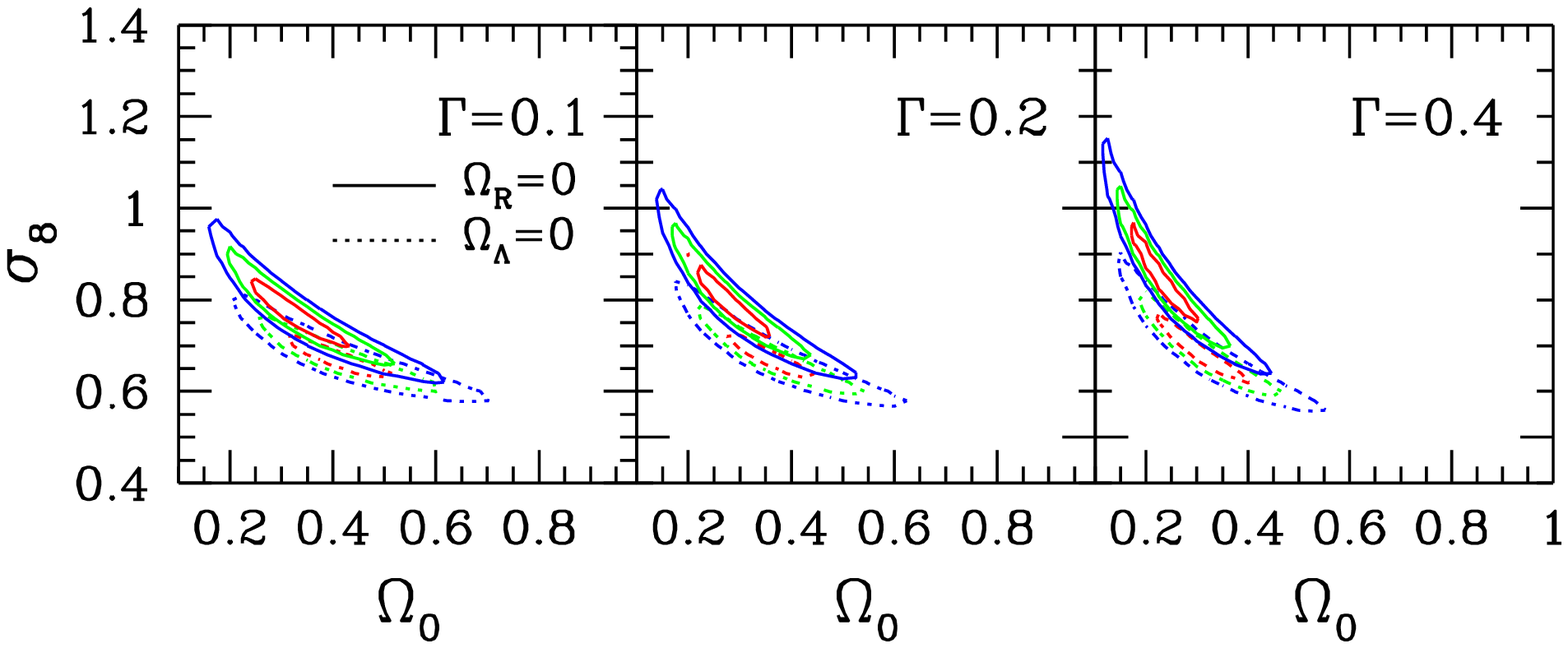}
\vspace{-7.3truecm}
\caption{Confidence regions on the $\Omega_m$--$\sigma_8$ plane.  In
all the panels, solid contours and dashed contours are for flat and
open models, respectively.  Here $\alpha=3.5$, $A=0$ and $\beta=1.15$
are assumed for the mass--luminosity conversion. Contours
are $1\sigma$, $2\sigma$ and $3\sigma$ c.l. for two significant
parameters.}
\label{fi:lz_like_gam}
\end{figure}

\section{$X$--ray cluster bias: from luminosity to mass}
The Press-Schechter approach is used in our analysis, as it
provides an accurate mass function in the
range of masses probed by the RDCS \cite{BRTN99}.  The conversion from
masses to X-ray luminosities, which is required in analysis of any
flux-limited sample is implemented as follows: {\bf (a)} convert mass
into temperature by assuming virialization, hydrostatic equilibrium
and isothermal gas distribution; {\bf (b)} convert temperature into
bolometric luminosity according to $L_{bol}\propto T^{\alpha}(1+z)^A$;
{\bf (c)} compute the bolometric correction to the 0.5-2.0 keV band.

The critical step is represented by the choice for the
$L_{bol}$--$T_X$ relation.  Low redshift data for $T\magcir 3\,$keV
indicates that $\alpha \simeq 2.7$--3.5, depending on the sample and
the data analysis technique \cite{WJF}, with a reduction of the
scatter after account for the effect of cooling flows in central
cluster regions \cite{AE99}. At lower temperatures, evidence has been
found for a steepening of the $L_{bol}$--$T_X$ relation below 1 keV
\cite{Pon96}.  As for the evolution of the $L_{bol}$--$T_X$ relation,
existent data out to $z\simeq 0.4$ \cite{MS97} and, possibly, out to
$z\sim 0.8$ \cite{RdC99} are consistent with no evolution (i.e.,
$A\simeq 0$).  Instead of assuming a unique mass--luminosity
conversion, in the following we will show how final constraints on
cosmological parameters changes as the $L_{bol}$--$T_X$ and $M$--$T_X$
relations are varied.

\section{Analysis and results}
The RDCS subsample, that we will use in the following analysis, 
has a flux--limit of $S_{lim}=3.5\times 10^{-14}\fl$
and contains 81 clusters with measured redshifts out to $z=0.85$ over
a 33 sq. deg. area \cite{RosIAP}. 
In order to fully exploit the information provided by the RDCS, we
resort to a maximum--likelihood approach, in which model predictions
are compared to the RDCS cluster distribution on the $(L,z)$ plane.
To this purpose, let $\phi(L,z)$ be the Press--Schechter based
luminosity function, as predicted by a given model, so that 
$\phi(L,z)\,(dV/dz)$ $dz\,dL$ is the expected number density of
clusters in the comoving volume element $(dV/dz)\,dz$ and in the
luminosity interval $dL$. Therefore, the expected number of clusters
in RDCS lying in the $dz\,dL$ element of the $(L,z)$ plane is
$\lambda(z,L)dzdL=\rho(z,L)$ $f_{sky}[S(z,L)](dV/dz) dzdL$.  Here
$f_{sky}$ is the flux--dependent RDCS sky--coverage.

\begin{figure}
\epsfxsize 11.5truecm
\hspace{0.3truecm}\epsffile{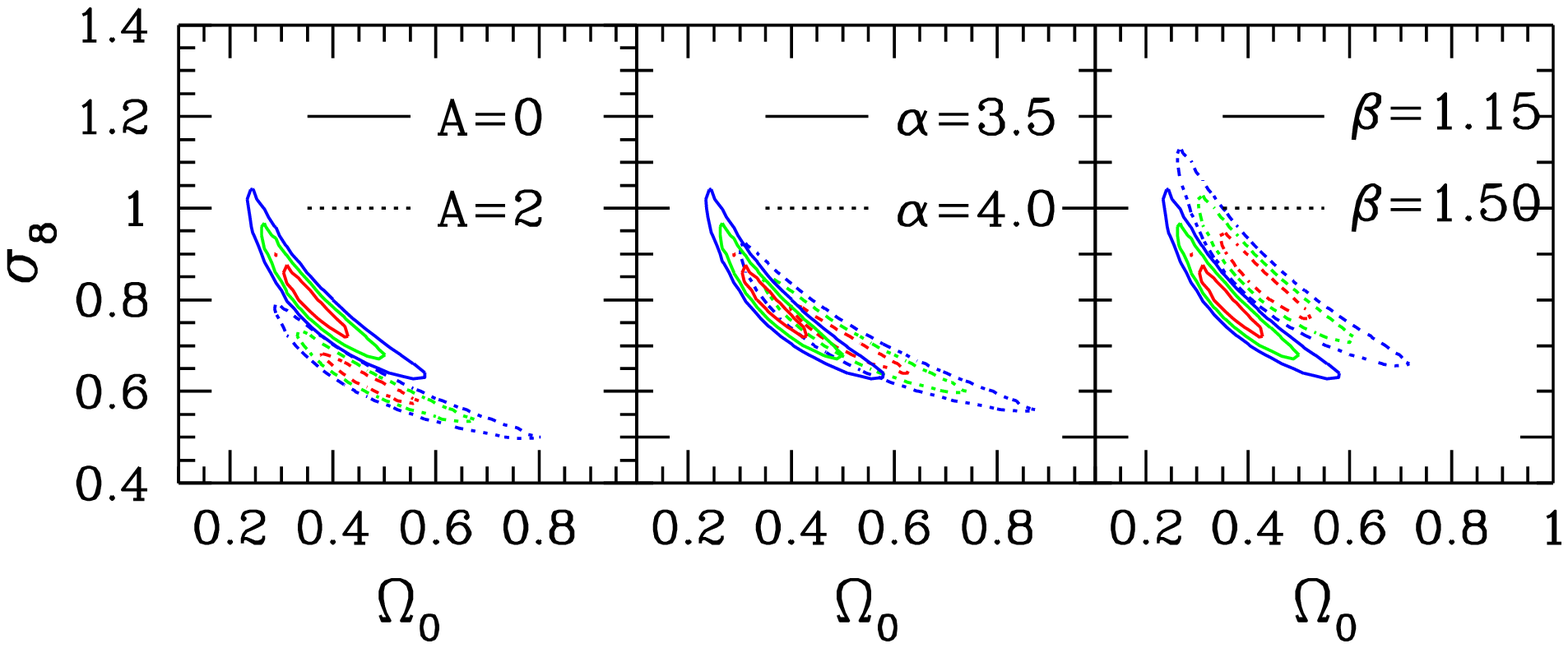}
\vspace{-7.3truecm}
\caption{Effect of changing the $L_{bol}$--$T_X$ relation.  Solid
contours are from assuming $\Gamma=0.2$, $\alpha=3.5$, $A=0$ and
$\beta=1.15$. Contours have the same meaning as in
Fig. \ref{fi:lz_like_gam}.}
\label{fi:lz_like}
\end{figure}

The likelihood function ${\cal L}$ is defined as the product of the
probabilities of observing exactly one cluster in $dz\,dL$ at each of
the $(z_i,L_i)$ positions occupied by the RDCS clusters, and of the
probabilities of observing zero clusters in all the other differential
elements of the $(z,L)$ plane which are accessible to RDCS. Assuming
Poisson statistics for such probabilities and defining $S=-2{\mbox{\rm
ln}}{\cal L}$, it is 
$S=-2\sum_{i=1}^{N_{occ}}{\mbox{\rm
ln}}[\rho(z_i,L_i)]+2\int dz \int dL\,\lambda(z,L)$, 
where the sum runs over the occupied elements of the $(z,L)$ plane.
Model predictions are also convolved with statistical errors on
measured fluxes, as well as with uncertainties in the luminosity--mass
relation associated to a $\simeq 30\%$ scatter in the $L_{bol}$--$T_X$
relation and to a $20\%$ uncertainty in the mass--temperature
conversion. Best estimates of the model parameters are obtained by
minimizing $S$.

In Figure \ref{fi:lz_like_gam} we show the resulting constraints on
the $\sigma_8$--$\Omega_m$ plane for different values of the shape
parameter $\Gamma$, based on assuming $\alpha=3.5$ and $A=0$ for the
$L_{bol}$--$T_X$ relation.  It is clear that low--density models are
always preferred, quite independent of $\Gamma$. We find
$\Omega_m=0.35^{+0.35}_{-0.25}$ and $\sigma_8=0.76^{+0.38}_{-0.14}$
($\Omega_m=0.42^{+0.35}_{-0.27}$ and $\sigma_8=0.68^{+0.21}_{-0.12}$)
for flat (open) models, where uncertainties correspond to $3\sigma$
confidence level for three significant fitting parameter.  No
significant constraints are instead found for $\Gamma$.
In order to verify under which circumstances a critical density model
may still be viable, we show in Figure \ref{fi:lz_like} the effect of
changing the parameters of the $L_{bol}$--$T_X$ relation. Although
best--fitting values of $\Omega_m$ and $\sigma_8$ move somewhat on the
parameter space, neither a rather strong evolution nor
a quite steep profile for the $L_{bol}$--$T_X$ relation can accommodate
a critical density Universe: an $\Omega_m=1$ Universe is always a
$>3\sigma$ event, even allowing for values of the $A$ and $\alpha$
parameters which are strongly disfavored by present data.

Based on these results, we point out that deep flux--limited $X$--ray
cluster samples, like RDCS, which cover a large redshift baseline
($0.1\mincir z\mincir 1.2$) and include a fairly large number of
clusters ($\magcir 100$) do indeed place significant constraints on
cosmological models. To this aim, some knowledge of the
$L_{bol}$--$T_X$ evolution is needed from a (not necessarily complete)
sample of distant clusters out to $z\sim 1$.

%

%
\vfill

\begin{iapbib}{99}{

\bibitem{AE99} Arnaud, K.A., \& Evrard, A.E. 1999, MNRAS, 305, 631
\bibitem{Bah98} Bahcall, N.A., \& Fan, X. 1998, ApJ, 504, 1  
\bibitem{BRTN99} Borgani, S., Rosati, P., Tozzi, P., \& Norman, C. 1999,
ApJ, 517, 40
\bibitem{RdC99} Della Ceca, R., et al. 1999, A\&A, in press, astro-ph/9910489
\bibitem{Don99} Donahue, M., \& Voit, G.M. 1999, ApJ, 523, L127
\bibitem{ECFH} Eke, V.R., Cole, S., Frenk, C.S., \& Henry, J.P. 1998, MNRAS,
298, 114
\bibitem{Geal98} Girardi, M., et al. 1998, ApJ, 506, 45
\bibitem{MS97} Mushotzky, R.F., \& Scharf, C.A. 1997, ApJ, 482, L13
\bibitem{Pon96} Ponman, T.J., et al. 1996, MNRAS, 283, 690
\bibitem{Real99} Reichart, D.E., et al. 1998, ApJ, 518, 521
\bibitem{RosIAP} Rosati, P. 1998, in Wide Field Surveys in Cosmology,
14th IAP Meeting (Paris, Publ.: Editions Frontieres) p.219
\bibitem{RDCS} Rosati, P., et al. 1998, ApJ, 492, L21
\bibitem{SBO98} Sadat, R., Blanchard, A., \& Oukbir, J. 1998, A\&A, 329, 21
\bibitem{VL99} Viana, P.T.P., \& Liddle, A.R. 1999, MNRAS, 303, 535
\bibitem{WJF} White, D.A., Jones, C., \& Forman, W. 1997, MNRAS, 292, 419
}
\end{iapbib}
\end{document}